# Nonlinear Compression of Besselon Waves for High Repetition-Rate Subpicosecond Pulses Trains

Anastasiia Sheveleva and Christophe Finot

*Abstract*—We theoretically and experimentally demonstrate the generation of high-quality low duty-cycle pulse trains at repetition rates of 28 GHz, 56 GHz and 112 GHz. Starting from a continuous wave we benefit from phase modulations in the temporal and spectral domains by applying a sinusoidal profile and a set of well-chosen π shifts, respectively, to generate a train of modified besselons at doubled repetition rate. With further nonlinear spectral expansion in a normally dispersive fiber followed by dispersion compensation we achieve subpicosecond durations and a duty cycle as low as 0.025 at 28 GHz. Spectral cancelation of one component over two or four enables to further double or quadruple the repetition rate.

## I. INTRODUCTION

The need of picosecond pulses at repetition rates of several tens of GHz has stimulated many fundamental and applied researches in the photonics community. Solutions such as mode-locked fiber lasers have been proposed [1] but remain expensive and require fine tuning of the cavity. Benefiting of the recent progress of high-quality microresonators is another very promising approach: tailoring the frequency comb generated in these microcavities has successfully provided pulse trains at repetition rates from a few tens up to hundreds of GHz. However, the repetition rate of the source is fully determined by the component design and cannot be continuously adjusted. Moreover, the coupling of the initial seed laser with the very narrow resonances of the resonators is still technologically challenging.

As a consequence, cavity-free designs, where pulse train properties directly depend on an external RF modulation, remain an attractive and versatile option. Different schemes are in this context possible. For example, one way is to apply a sinusoidal intensity modulation on a continuous wave laser then followed by a nonlinear compression in one or several segments of fibers [2, 3]. High quality pulse trains are generated but achieving very low duty cycles requires an increasingly complex architecture. A second possible scheme relies on imprinting a temporal sinusoidal phase modulation combined with propagation in a linearly dispersive medium [4, 5]. Ultrashort structures have been demonstrated, but the pulse quality is severely impaired by strong sidelobes or continuous background. Several nonlinear methods, including nonlinear optical loop mirrors [6], additional synchronized amplitude modulation [7] or Mamyshev-like devices [8] have been implemented to get rid of these spurious ripples but once again, they significantly increase the complexity of the setup.

Recently, we have highlighted numerically and experimentally that the results achieved with the second scheme can be significantly improved if the traditionally used quadratic spectral phase is replaced by a triangular one [9]. Fourier-transform limited waveforms are achieved and properly choosing the depth of the initial sinusoidal phase modulation provides an excellent extinction ratio. We named those structures besselons as their main features are governed by the properties of Bessel functions [10]. By applying an additional π phase shift on the central component, we have also demonstrated the generation of a modified besselon that has the advantage of being shorter while keeping an excellent pulse quality. Moreover, it can be time-multiplexed to efficiently double the repetition rate. However, this linear processing only gives structures with a limited duty-cycle of 0.201.

In the present contribution, we combine the modified besselon with nonlinear compression techniques to demonstrate a decrease of the duty cycle by nearly one order of magnitude. Starting from a 14 GHz RF phase modulation, we successfully experimentally demonstrate subpicosecond pulses at repetition rates up to 112 GHz.

## II. PRINCIPLE OF OPERATION

The principle of our proposed architecture is illustrated in Fig. 1. Starting from continuous wave (CW) which phase has been modulated in the temporal domain by a sinusoidal profile with a depth of modulation $A_m$ and an angular frequency $\omega_m$, we apply a triangular spectral profile made of π/2 discrete phase shifts combined with an additional central π phase shift (refer to the blue line in Fig. 2(b)). Such a processing leads to the

We acknowledge the support of the Institut Universitaire de France (IUF), the Bourgogne-Franche Comté Region. The experiments have benefited from the PICASSO experimental platform of the University of Burgundy. We thank Bertrand Kibler and Sonia Boscolo for fruitful discussions as well as Ugo Andral for initial developments of the Besselon experimental concept.

A. Sheveleva and C. Finot, are with the Laboratoire Interdisciplinaire Carnot de Bourgogne, UMR 6303 CNRS-Université Bourgogne-Franche-Comté, 9 avenue Alain Savary, BP 47870, 21078 Dijon Cedex, France (e-mail: christophe.finot@u-bourgogne.fr).

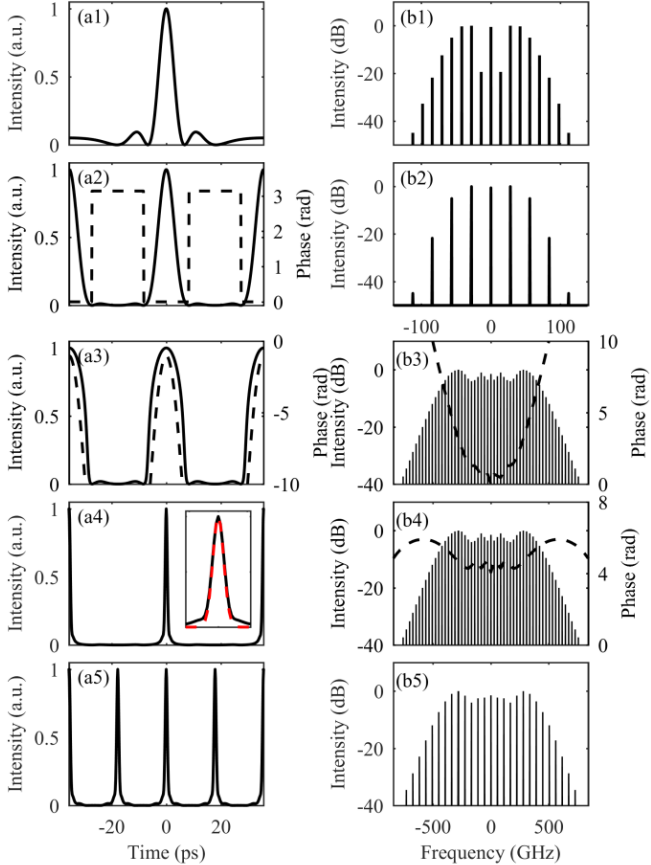

Fig. 1. Numerical simulation of the evolution of the (a) temporal and (b) spectral properties of the shaped signal at different stages. Full and dashed lines are used for the intensity profiles and the phase profiles, respectively. Modified besselon arising from the sinusoidal phase modulation with a depth of $A_m = 3.72$ rad and frequency 14 GHz followed by the spectral phase cancelation is depicted in panels (1). Panels (2) display profiles at the doubled repetition rate of 28 GHz after cancellation of the odd spectral components. The properties of the wave after the nonlinear propagation before and after applying a quadratic spectral phase are shown in panels (3) and (4), respectively. Inset of Fig. 4(a) is a magnification of the central part of the pulse, which is compared to a Gaussian fit. Panels (5) show a case of further doubling of the repetition rate reaching 56 GHz.

generation of a coherent structure we named optical modified besselon. In ref [10], we provided an in-depth analytical analysis of this new waveform which temporal profile is analytically expressed as (for $A_m > 2.4$ rad):

$$\psi'_B(t) = -J_0(A_m) + 2\sum_{n=1}^{\infty} J_n(A_m) \cos(n\,\omega_m\,t) \qquad (1)$$

where $J_n$ are the Bessel functions of the first kind of order $n$. The resulting temporal intensity profile (see panel (a1) in Fig. 1) highlights that the initial phase modulated signal reshapes into a train of short structures at the repetition rate $\omega_m$. However, the pulses are impaired by strong sidelobes and high background. Interestingly, we have shown in ref. [10] that a very efficient way to get rid of these spurious ripples while doubling the repetition rate is to cancel the odd frequency components of spectrum. The resulting field is:

$$\psi'_{2B}(t) = \cos\big(A_m \sin(\omega_m t)\big) - 2 J_0(A_m). \qquad (2)$$

and presents a duty cycle of 0.201 with an optimum signal extinction ratio obtained for $A_m = 3.72$ rad.

In order to further decrease the pulse duration, we benefit from nonlinear spectral expansion by combining the initial linear stage devoted to generation of the modified besselon with propagation in a highly nonlinear fiber (HNLF). In contrast to soliton-like compression [11] or to Mamyshev-like schemes [8], we have found that compression based on normally dispersive fibers [12] leads to the best performance in terms of pulse quality (level of resulting sidelobes, extinction ratio, symmetry). We consider here propagation in a HNLF with parameters corresponding to the experimental demonstration described in the next section: nonlinear coefficient $\gamma$ of 10 /W/km, second order dispersion coefficient $\beta_2$ of 0.89 ps$^2$/km, attenuation of 0.5 dB/km. The evolution of light in the HNLF can be predicted using the well-known nonlinear Schrodinger equation solved by the split-step Fourier algorithm. We notice in panel (b3) that, for an average input power of 27 dBm, the spectrum has been significantly expanded to reach a width of 1.4 THz. Oscillations of the spectrum are rather moderate. Combination of Kerr nonlinearity with normal dispersion leads in the temporal domain to the reshaping of the besselon pulses towards a parabolic like structure having a parabolic phase. The pulses have broadened and care should be devoted to avoid overlapping of the neighboring pulses that may ultimately lead to the generation of dark structures in the overlapping region. Temporal compression of the parabolic structures can be achieved by applying a quadratic spectral phase [12]. The resulting profiles, shown in panel (a4), stress the quality of the resulting pulse train. Durations as short as 0.95 ps are achieved, leading to a duty cycle as low as 0.034. Quite remarkably, an excellent extinction ratio is maintained and no spurious sidelobes are visible. Consequently, it is possible to temporally interleave those pulses to further increase the repetition rate. For example, suppressing one frequency component over two leads to the results plotted in panel (a5) with an excellent quality maintained. Note than an alternative technological solution could be the use of the fractional Talbot effect [13].

### III. EXPERIMENTAL SETUP

The experimental setup we implemented is sketched in Fig. 2(a) and is based on devices typical of the telecommunication industry. The number of optical elements is voluntarily limited. An initial CW at 1550 nm is delivered by an external cavity laser. In order to generate the modified besselon, the temporal phase of the CW is modulated by a sinusoidal RF signal delivered by an electrical clock running at a frequency of 14 GHz. The modulation depth of 3.72 rad which is required to generate the optimum waveform is achieved thanks to a low-$V_\pi$ lithium niobate phase modulator (PM). The spectral phase is then sculptured using a linear spectral shaper [14]. To double the repetition rate and to simultaneously make the waveform Fourier transform limited, we use the phase only spectral mask plotted in red in subplot (b) of Fig. 2. Rather than using the direct programming amplitude attenuation, which would have been strongly impacted by the resolution of our shaper, we have imprinted additional $\pi/2$ phase shifts on the odd components, so that the resulting $\pi$ phase difference translates into notch filters (compare the red line with the unmodified phase profile in blue). The level of attenuation of an amplified spontaneous emission (Fig. 2(b), black line) confirms that a narrow bandwidth attenuation higher than 20 dB is achieved.

The frequency doubled signal is then amplified by an erbium-doped fiber amplifier (EDFA) that delivers up to 27 dBm of average power. The nonlinear spectral expansion takes place in a normally dispersive HNLF with properties reported in the previous section. The high level of initial phase modulation enables us to avoid the deleterious effects of Brillouin backscattering. Contrary to other cavity-free schemes based on nonlinear sculpturing, it is therefore not required to insert an additional phase modulator aimed at Brillouin mitigation that could induce some additional jitter [3]. At the output of the nonlinear fiber, a second programmable filter is used to compensate the chirp of the signal by imprinting a quadratic spectral phase [14] (see spectral profile in Fig. 2(c1)). Note that such phase compensation could also have been achieved using fiber Bragg gratings or a simple piece of anomalous fiber operating in the linear regime of propagation. The spectral shaper also enables us to increase the repetition rate by canceling one component over two or over four on the nonlinearly broadened spectrum (see attenuation masks in Fig. 2 (c2-3)). Imprinting attenuation also helps remove partly the amplified spontaneous emission introduced by the EDFA.

The detection of the temporal properties of the shaped pulses has been achieved thanks to an optical sampling oscilloscope (OSO). For measurement of the temporal durations below the resolution of the OSO (1 ps), we also involved an autocorrelator based on second-harmonic generation. Spectral properties are recorded on a high-resolution optical spectrum analyzer that was also used for the fine adjustment of the linear shaper.

## IV. EXPERIMENTAL VALIDATION

### A. Generation of pulse trains at 14 and 28 GHz

The experimental results achieved for a repetition rate of 14 and 28 GHz are summarized in Fig. 3. After the stage of initial temporal phase modulation and spectral phase cancelation (panels 1), the temporal and spectral intensity profiles are found in perfect agreement with the analytical predictions based on Eq. (1). The phase modulation combined of discrete $\pi$ phase shifts efficiently kills the odd spectral components and increases the repetition rate up to 28 GHz where the experimental temporal and spectral properties are once again found in excellent agreement with the theoretical predictions of Eq. (2). The extinction ratio is excellent and the FWHM duration of the resulting pulses is 7.2 ps, leading to a duty cycle of 0.2. Panel (b3) highlights the nonlinear signal expansion occurring in the HNLF for an input average power of 27 dBm. The spectrum has broadened symmetrically and presents a width at -20dB that has been increased by a factor 6.25 and becomes larger than 1.4 THz. The temporal duration has increased up to 9.2 ps. The combination of normal dispersion and Kerr nonlinearity has led to the expected parabolic-like temporal intensity profile [3]. A parabolic spectral phase applied on the second programmable filter compresses the pulse down to the picosecond level while maintaining a profile that is free from pronounced sidelobes. The pulse train measured on the optical sampling oscilloscope is plotted on panel (a4). It has a duration of 1.12 ps that is comparable with the resolution of the device. Measurements made in the persistent mode are provided on top of the panel (a4) and stress the high stability of the pulse train. In order not to be impaired by the temporal resolution of the OSO, we also recorded the autocorrelation signal. Results are plotted in panel (c) and confirm the absence of significant pedestals. Comparison between the experimental records and a Fourier transform limited waveform derived from the optical spectrum highlights that the quadratic phase compensation has led to a nearly ideal waveform. Fitting with a Gaussian waveform indicates a temporal duration close to 0.89 ps, leading to a duty-cycle of 0.025.

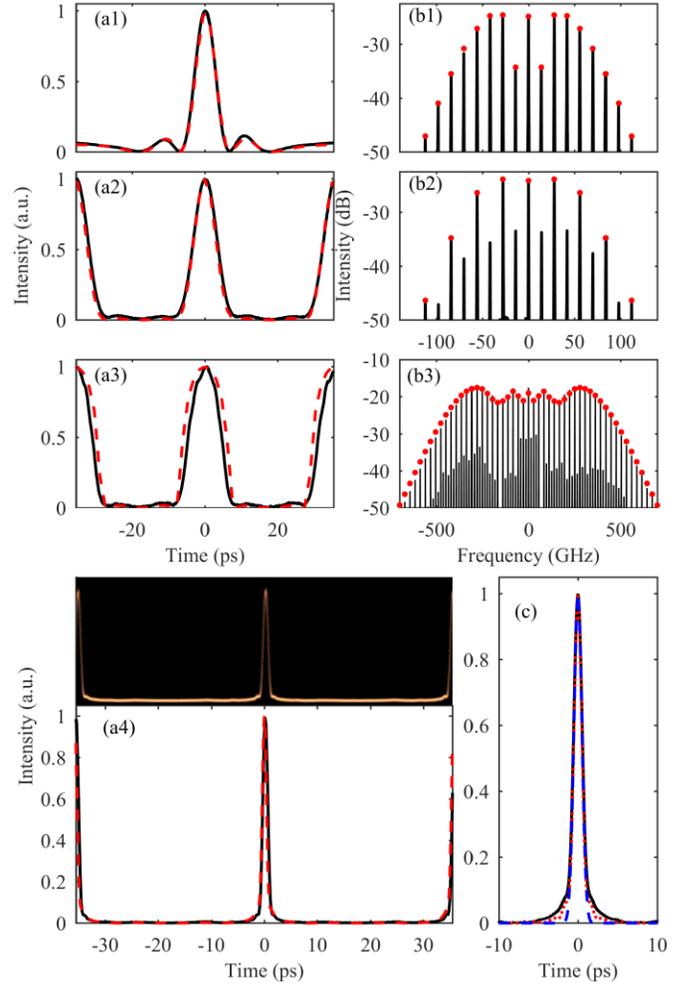

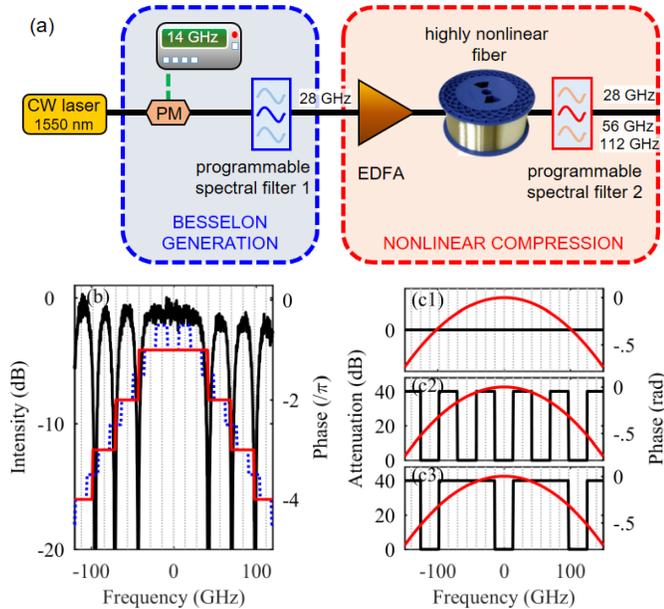

Fig. 2. (a) Experimental setup. (b) Phase profile used to obtain the modified besselon at 14 GHz is depicted in blue, while red line shows phase mask with additional shifts used to double the repetition rate. Amplified spontaneous emission (in black) modulated with the latter phase profile stresses appearance of the notch filters. (c) Spectral phase and attenuation profiles programmed on the second programmable filter (red and black lines respectively). Profiles for pulse trains at 28 GHz, 56 GHz and 112 GHz are plotted of panels (c1), (c2) and (c3) respectively.

Fig. 3. (a) Temporal and (b) spectral properties of the pulse train recorded at different stages. Modified besselons achieved for $A_m$ = 3.72 rad at 14 GHz (panels 1) are compared with the results after repetition rate doubling (panels 2), and with the profiles achieved after nonlinear propagation (panels 3) and additional quadratic spectral phase mitigation (panels 4). Experimental records (black line) are compared with analytical predictions or numerical simulations (red). Top part of panel (a4) shows the temporal profiles recorded in the persistent mode of the optical sampling oscilloscope whereas panel (c) deals with the autocorrelation signal. The central autocorrelation peak is compared with the autocorrelation of a Gaussian pulse as well as the Fourier transform-limited waveform predicted from the optical spectrum.

*B. Generation of pulse train at 56 and 112 GHz*

Experimental results achieved when further doubling or quadrupling the repetition rate are summarized in Fig. 4. The various spectra (panels a) highlight the efficient suppression of one component over 2 or 4 while maintaining a high OSNR (higher than 40 dB when measured on a 5 MHz bandwidth). The temporal intensity profiles measured on the OSA confirm that a high quality is maintained. The temporal and amplitude jitters can be noticed which are mainly ascribed to the OSO device. Measurements made with the autocorrelator provide a FWHM duration of 0.89 ps and 1 ps (with the Gaussian assumption), leading to duty cycles of 0.05 and 0.11 respectively. The temporal measurements reveal that some bumps appear between the pulses when quadrupling the repetition rate up to 112 GHz. Such small bumps are reproduced by numerical simulations and indicate that higher repetition rates will be difficult to achieve.

## V. CONCLUSION

By using the nonlinear compression of modified besselon pulses, we have demonstrated the generation of sub-picosecond high quality pulses using a simple experimental architecture. Starting from an RF signal at 14 GHz, the combination of linear and nonlinear fiber-based processing has enabled to obtain high quality pulse trains with a duty cycle as low as 0.025 at 28 GHz. As the architecture is cavity-free, the repetition rate is fully tunable. Compared to the cavity-free solutions previously proposed [3, 6-8, 13], a single phase modulator is here required, less segments of fibers are involved and the absence of Brillouin backscattering avoids the implementation of a dithering modulation.

## REFERENCES

[1] T. F. Carruthers and I. N. Duling, "10-GHz, 1.3-ps erbium fiber laser employing soliton pulse shortening," *Opt. Lett.,* vol. 21, no. 23, pp. 1927-1929, 1996/12/01 1996.
[2] S. Pitois, C. Finot, J. Fatome, and G. Millot, "Generation of 20-Ghz picosecond pulse trains in the normal and anomalous dispersion regimes of optical fibers," *Opt. Commun.,* vol. 260, no. 1, pp. 301-306, 2006.
[3] I. El Mansouri, J. Fatome, C. Finot, M. Lintz, and S. Pitois, "All-Fibered High-Quality Stable 20- and 40-GHz Picosecond Pulse Generators for 160-Gb/s OTDM Applications," *IEEE Photon. Technol. Lett.,* vol. 23, no. 20, pp. 1487-1489, 2011.
[4] T. Kobayashi, H. Yao, K. Amano, Y. Fukushima, A. Morimoto, and T. Sueta, "Optical pulse compression using high-frequency electrooptic phase modulation," *IEEE J. Quantum Electron.,* vol. 24, no. 2, pp. 382-387, 1988.
[5] T. Komukai, Y. Yamamoto, and S. Kawanishi, "Optical pulse generator using phase modulator and linearly chirped fiber Bragg gratings," *IEEE Photon. Technol. Lett.,* vol. 17, no. 8, pp. 1746-1748, 2005.
[6] S. Yang and X. Bao, "Generating a high-extinction-ratio pulse from a phase-modulated optical signal with a dispersion-imbalanced nonlinear loop mirror," *Opt. Lett,* vol. 31, no. 8, pp. 1032-1034, 2006/04/15 2006.
[7] H. Hu *et al.*, "Pulse source based on directly modulated laser and phase modulator," *Opt. Express,* vol. 15, no. 14, pp. 8931-8937, 2007/07/09 2007.
[8] D. Wang *et al.*, "Pedestal-free 25-GHz subpicosecond optical pulse source for 16 x 25-Gb/s OTDM based on phase modulation and dual-stage nonlinear compression," *Appl. Opt.,* vol. 57, no. 11, pp. 2930-2934, 2018/04/10 2018.
[9] U. Andral, J. Fatome, B. Kibler, and C. Finot, "Triangular spectral phase tailoring for the generation of high-quality picosecond pulse trains," *Opt. Lett.,* vol. 44, no. 19, p. 4913, 2019.
[10] A. Sheveleva, U. Andral, B. Kibler, S. Boscolo, and C. Finot, "Temporal Optical Besselon Waves for High-Repetition Rate Picosecond Sources," p. arXiv:2003.12630, 2020.
[11] L. F. Mollenauer, R. H. Stolen, and J. P. Gordon, "Experimental observation of picosecond pulse narrowing and solitons in optical fibers," *Phys. Rev. Lett.,* vol. 45, no. 13, pp. 1095-1098, 1980.
[12] C. V. Shank, R. L. Fork, R. Yen, R. H. Stolen, and W. J. Tomlinson, "Compression of femtosecond optical pulses," *Appl. Phys. Lett.,* vol. 40, pp. 761-763, 1982.
[13] D. Marion and J. Lhermite, "Electro-optically modulated lasers: pulse energy and contrast enhancement derived from theoretical spectrum," *Opt. Lett.,* vol. 45, no. 9, pp. 2664-2667, 2020.
[14] M. A. F. Roelens *et al.*, "Dispersion trimming in a reconfigurable wavelength selective switch," *J. Lightw. Technol.,* vol. 26, pp. 73-78, 2008.

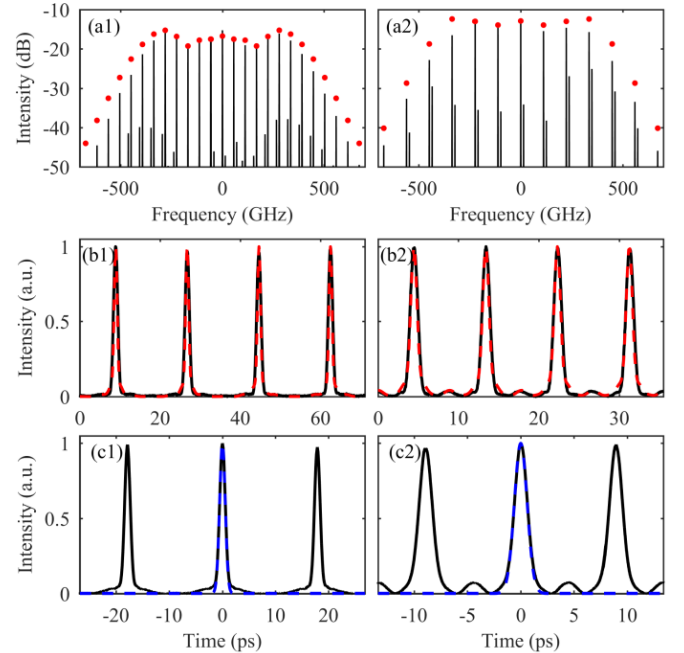

Fig. 4 Properties of the signal after additional doubling and quadrupuling of the repetition rate reaching 56 and 112 GHz are displayed in panels (1) and (2), respectively. (a) Optical spectra. (b) Temporal intensity profiles recorded on the optical sampling oscilloscopes. (c) Autocorrelation signals (black) fitted by the autocorrelation of a Gaussian (blue).